\documentclass[prb,aps,twocolumn,showpacs,amsmath,amssymb,citeautoscript]{revtex4}
\usepackage{graphicx}
\usepackage{dcolumn}
\usepackage{bm}
\usepackage{times}
\usepackage{amsmath}
\usepackage{amssymb}
\usepackage[normal]{subfigure}

\usepackage{color}

\makeatletter

\begin{document}

\title{Detecting a Majorana-Fermion Zero Mode Using a Quantum Dot}

\author{Dong E. Liu, and Harold U. Baranger}
\affiliation{
Department of Physics, Duke University, Box 90305, 
Durham, North Carolina 27708-0305, USA
}

%\date{July 21, 2011}
\date{\today}

\begin{abstract}
We propose an experimental setup for detecting a Majorana zero mode consisting of 
a spinless quantum dot coupled to the end of a \emph{p}-wave superconducting nanowire. 
The Majorana bound state at the end of the wire strongly influences the conductance through the quantum dot: driving the wire through the topological phase transition causes a sharp jump in the conductance by a factor of $1/2$.
In the topological phase, the zero temperature peak value of the dot conductance (i.e.\ when the dot is on resonance and symmetrically coupled to the leads) is $e^2/2 h$. 
In contrast, if the wire is in its trivial phase, the conductance 
 peak value is $e^2/h$, or if a 
regular fermionic zero mode occurs on the end of the wire, the conductance is $0$. 
The system can also be used to tune Flensberg's qubit system [PRL 106, 090503 (2011)] to the required degeneracy point.
\end{abstract}

\pacs{73.21.-b, 74.78.Na, 73.63.-b, 03.67.Lx}
%71.10.Pm Fermions in reduced dimensions (anyons, composite fermions,
%Luttinger liquid, etc.) (for anyon mechanism in superconductors, see
%74.20.Mn)
%74.20.Mn   Nonconventional mechanisms 
%    HUB: ABOVE NOT APPROPRIATE!! 74.20 is for theories and models 
%          of the superconducting state-- not what we're doing! 
%74.78.Na   Mesoscopic and nanoscale systems   HUB: OK
%73.63.-b   Electronic transport in nanoscale materials and structures  HUB: OK
%03.67.Lx   Quantum computation architectures and implementations  HUB: OK
%
%74.78.-w		Superconducting films and low-dimensional structures
%74.78.Fk		Multilayers, superlattices, heterostructures
%74.78.Na		Mesoscopic and nanoscale systems
%
%73.21.-b		Electron states and collective excitations in multilayers, quantum wells, mesoscopic, and nanoscale systems
%

\maketitle

Majorana fermions, an exotic type of quasi-particle with non-Abelian statistics, are attracting a great deal of attention due to both their fundamental interest and their potential application for decoherence-free quantum computation. 
%Several schemes for realizing 
Several ways to realize unpaired Majorana fermions in a vortex core in a \emph{p}-wave superconductor \cite{moore91,read00,fu08,sato09,sau10,aliceaPRB10} and superfluid \cite{kopnin91,tewari07} have been proposed. Majorana bound states (MBS) may also be realized at the ends of a one-dimensional \emph{p}-wave superconductor \cite{kitaev01} for which the proposed 
%physical 
system is a semiconductor nanowire with Rashba spin-orbit interaction to which both a magnetic field and proximity-induced \emph{s}-wave pairing 
are added \cite{lutchynPRL10,oreg10}. In view of these proposals, how to detect and verify the existence of MBS becomes a key issue. 
Suggestions include noise measurements \cite{bolech07,nilsson08}, resonant Andreev reflection by an STM \cite{law09}, and $4\pi$ periodic 
Majorana-Josephson currents \cite{kitaev01,fuPRB09,lutchynPRL10,oreg10}.

With regard to quantum computation, the braiding of Majorana bound states in a network of wires by applying a ``keyboard'' of individually tunable gates \cite{alicea11} leads to non-trivial computation. 
Note that all the detecting methods proposed to date \cite{kitaev01,bolech07,nilsson08,fuPRB09,law09,lutchynPRL10,oreg10}, involving electron transfer into or out of MBS, will destroy the qubit information. 
In addition, such braiding can not result in universal quantum computation; it must be supplemented by a topologically 
unprotected $\pi/8$ phase gate \cite{boykin99}. Recently, Flensberg introduced a system consisting of a quantum dot coupled to 
two MBS (MBS-dot-MBS) through which this $\pi/8$ phase gate can be achieved \cite{flensberg11}. A key point is that the system must be 
fine tuned so that the ground state is degenerate \cite{flensberg11}. 
%In that sense, probing the MBS signature by using quantum dot is necessary and interesting, which may provide away to tune Flensberg's qubit to energy degenerate point. HUB: can't put this sentence in yet because you haven't introduced the dot!!

\begin{figure}[t]
\centering
\includegraphics[width=3.3in,clip]{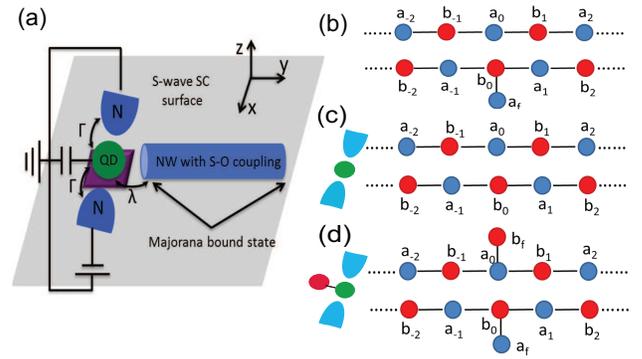}
\caption{(color online) (a) Sketch of dot-MBS syste:the semiconductor wire on a \emph{s}-wave superconductor surface, 
and a magnetic field perpendicular to the surface ($\hat{z}$ direction). The dot couples to one end of the wire; 
the conductance through the dot is measured by adding two external leads. (b) Majorana chain representation for 
leads-dot-MBS system ($G_{\rm peak}=e^2/2h$). (c) Dot-leads system with nothing side-coupled (left) and Majorana chain 
representation (right) ($G_{\rm peak}=e^2/h$). (d) Dot-leads system with side-coupled regular fermionic zero mode (left) 
and Majorana chain representation (right) ($G_{\rm peak}=0$). 
}
\label{fig:setup}
\end{figure}

In this work, we consider a spinless quantum dot coupled to a MBS at the end of a \emph{p}-wave superconducting (SC) nanowire, 
and study the conductance, $G$, through the dot by adding two external leads (schematic in Fig.\,\ref{fig:setup}). 
We find that the conductance is independent of the properties of the MBS, the nanowire, or the superconductor. 
The dependence of $G$ on the dot properties has the same functional form whether an MBS is present or not. Therefore, the 
conductance behavior can be conveniently summarized by its peak value, when the dot is on resonance and symmetrically 
coupled to the probing leads. It is $e^2/2h$ in the topological SC phase, $G_{\rm peak} = 1/2$, in contrast to that for a 
dot coupled to a regular fermionic zero mode, $G_{\rm peak} = 0$, as well as to that for a dot coupled to the wire in its 
topologically trivial phase, $G_{\rm peak} = 1$. Thus, as the wire is driven through the topological phase transition, the conductance shows a sharp jump by a factor of $1/2$.
The conductance through the dot is, then, a probe of the presence of the MBS. Note that direct transfer between the MBS and dot is not necessary, 
though dephasing of the qubit is introduced when the dot is on-resonance. Such a ``less invasive'' sensing method provides a potential 
way to probe a MBS without totally destroying the information in the qubit. We also consider coupling the dot to both ends of the wire (two MBS), 
with a magnetic flux $\Phi$ through the loop. The conductance as a function of phase shows peaks at $\Phi=(2n+1)\pi\Phi_0$ which can 
be used to tune Flensberg's qubit system \cite{flensberg11} to the energy degeneracy point.

% We find that the on-resonance symmetric conductance {\color{red}(i.e. the peak value)} at zero temperature is $e^2/2h$ if the wire is in the topological SC phase, $G_{\rm res} = 1/2$. This result differs from that for a dot coupled to a regular fermionic zero mode ($G_{\rm res} = 0$), as well as from that for a dot coupled to the wire in its topologically trivial phase ($G_{\rm res} = 1$). {\color{red}As the wire is driven through the topological phase transition, the conductance shows a sharp sjump by a factor of $1/2$.} The conductance through the dot is, then, a probe of the presence of the MBS. Note that direct transfer between the MBS and dot is not necessary, though dephasing of the qubit is introduced when the dot is on-resonance. Such a ``less invasive'' sensing method provides a potential way to probe a MBS without totally destroying the information in the qubit. We also consider coupling the dot to both ends of the wire (two MBS), with a magnetic flux $\Phi$ through the loop. The conductance as a function of phase shows peaks at $\Phi=(2n+1)\pi\Phi_0$ which can be used to tune Flensberg's qubit system \cite{flensberg11} to the energy degeneracy point.

\begin{figure}[t]
\centering
\includegraphics[width=3.3in,clip]{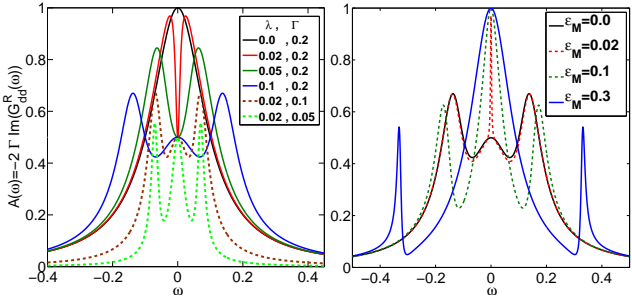}
\caption{(color online) Spectral function of the quantum dot in the on-resonance ($\epsilon_d=0$) and symmetric ($\Gamma_L=\Gamma_R=\Gamma/2$) case. (a) Coupling from dot to MBS ($\lambda$) and leads ($\Gamma$) varies at fixed $\epsilon_M=0$. Solid lines: $\Gamma=0.2$ and $\lambda$ from $0$ to $0.1$. Dashed lines: $\lambda=0.02$ and $\Gamma$ from $0.05$ to $0.1$. The spectral function evolves from a simple resonant tunneling form in the absence of coupling to a three-peak structure; the middle peak is a direct result of the Majorana zero mode. (b) MBS-MBS coupling strength varies at fixed $\Gamma=0.2$, $\lambda=0.1$. Note that $A(\omega=0)=1/2$ whenever a Majorana is coupled. The unit is chosen so that the lead band width is $D_L=40$ for all calculations. 
}
\label{fig:MBS_spectral}
\end{figure}

{\em Single MBS---}We consider the setup shown in Fig.\,\ref{fig:setup}(a) in which a spinless quantum dot is coupled 
to the end of a semiconductor nanowire with strong Rashba spin-orbit interaction, proximity-induced \emph{s}-wave 
superconductivity, and a magnetic field $B$ \cite{lutchynPRL10,oreg10}. We assume the nanowire and superconductor 
are not grounded and have a negligible charging energy. The magnetic field is smaller than the superconductor's upper 
critical field, but the Zeeman splitting $V_{z}=g \mu_B B / 2$ must be large enough for the wire to be in the topological 
SC phase, $V_{z}>\sqrt{\Delta^2+\mu^2}$ where $\Delta$ is the SC order parameter and $\mu$ is the chemical potential of the wire. 
Isolated Majorana fermion zero modes $\eta_1$ and $\eta_2$ appear in this case at the two ends of the wire. Suppose the dot is coupled 
to $\eta_1$ and the operators $d^{\dagger}$ ($c_{k\alpha}^{\dagger}$) create an electron in the dot (leads). 
The Hamiltonian can then be written as \cite{numberraiselower}
\begin{equation}
H=H_{\textrm{Leads}}+H_{\textrm{Dot}}+H_{\textrm{D-L}}+i\epsilon_{M}\eta_{1}\eta_{2}+\lambda(d-d^{\dagger})\eta_{1} ,
\label{eq:H_DM}
\end{equation}
where $H_{\textrm{Leads}} \!=\! \sum_{k}\sum_{\alpha=L,R}\epsilon_{k}c_{k\alpha}^{\dagger}c_{k\alpha}$ describes the left and right metallic leads with chemical potential $\mu_\textrm{lead} \!=\! 0$, 
$H_{\textrm{Dot}} \!=$ $\epsilon_{d}d^{\dagger}d$ describes the dot with a gate tunable level $\epsilon_d$, and
$H_{\textrm{D-L}} \!=\! \sum_{\alpha=L,R}\sum_{k}V_{\alpha}(c_{k\alpha}^{\dagger}d+h.c.)$
describes the coupling between the dot and the leads. $\epsilon_{M}\sim e^{-L/\xi}$ is the coupling between the two Majorana bound states, where $L$ is the length of the wire and $\xi$ is the superconducting coherence length. 

The last part of $H$ describes the coupling between the dot and MBS. Here, we assume that the Zeeman splitting is the largest scale $V_z\gg |V_{\rm{bias}}|, T, \Gamma, \lambda$, where $V_{\rm{bias}}$ is the source-drain voltage, $T$ is temperature, and $\Gamma=\Gamma_L+\Gamma_R$ is the dot-leads coupling with $\Gamma_{\alpha} \equiv \pi |V_{\alpha}|^2 \rho_0$ and $\rho_0$ the density of states of the leads. In this case, one need only consider a spinless single level in the dot. It is helpful to switch from the Majorana fermion representation to the completely equivalent regular fermion one by defining $\eta_{1}=\left(f+f^{\dagger}\right)/\sqrt{2}$ and $\eta_{2}=i\,\left(f-f^{\dagger}\right)/\sqrt{2}$. The last two terms in $H$ become
\begin{equation}
H_{\textrm{MBS}}=\epsilon_{M}(f^{\dagger}f-\frac{1}{2})+\lambda(d-d^{\dagger})\left(f+f^{\dagger}\right)/\sqrt{2} \;.
\end{equation}

The linear conductance through the lead/dot/lead system is related to the Green function of the dot level, $G_{dd}^{R}(\omega)$, by 
\begin{equation}
\label{KuboFormular}
G=\frac{e^{2}}{h} \!\int\! \frac{d\omega}{2\pi}\frac{\Gamma_{L}\Gamma_{R}}{\Gamma_{L}+\Gamma_{R}}\left(-2 
\operatorname{Im}\left[G_{dd}^{R}(\omega)\right]\right)\left(-\frac{\partial n_{f}}{\partial\omega}\right).
\end{equation}
The standard equation of motion method yields an exact expression for the Green function \cite{Leijnse10},
\begin{equation}
G_{dd}^{R}(\omega)=\frac{1}{\omega-\epsilon_{d}+i \Gamma-|\lambda|^{2}K(\omega)[1+|\lambda|^{2}\tilde{K}(\omega)]} \;,
\end{equation}
with $K(\omega)=1/(\omega-\epsilon_{M}^{2}/\omega)$
and 
\begin{equation}
\tilde{K}(\omega)=\frac{K(\omega)}{\omega+\epsilon_{d}+i \Gamma-|\lambda|^{2}K(\omega)} \;.
\end{equation}
%$G_{dd}^{R}(\omega)$ was also obtained \cite{Leijnse10} for other purposes.
%HUB: I took out above line and just cited it above. Is it OK?
For $\epsilon_M=0$ and $\epsilon_d=0$, one has $G_{dd}^{R}(\omega\rightarrow 0) =1 / 2(\omega+i\Gamma)$, 
and so the on-resonance ($\epsilon_d=0$) and symmetric ($V_L=V_R$),i.e. peak, conductance at zero temperature is
\begin{equation}
 G_\textrm{peak} =-(e^{2}/h)\,\Gamma \operatorname{Im}[G_{dd}^{R}(\omega\rightarrow0)]
 = e^{2}/2h \;.
\end{equation}
This result is distinct from both the case of a dot coupled to a regular fermionic zero mode, which gives $G_\textrm{peak} \!=\! 0$ \cite{FerryGoodnickBird}, and that of a dot disconnected from the wire, for which $G_\textrm{peak} \!=\! e^2/h$.
For asymmetric coupling ($V_L \!\neq\! V_R$), there is a pre-factor $4\Gamma_L \Gamma_R/(\Gamma_L + \Gamma_R)^2$ for all cases. \textit{Therefore, the signature of the Majorana fermion is that the conductance is reduced by a factor of $1/2$.}
%, to $G=e^2/2h$ in the symmetric case.  
%Therefore, the result $G=e^2/2h$ is a signature of Majorana fermion. 

To further understand this result, we rewrite the model in the Majorana representation \cite{kitaev01}. The probe leads are described by two semi-infinite tight-binding fermionic chains ${c_i}$ ($i \!=\! ...,-1, 0, 1, 2, ...$) joined at the dot, $i \!=\! 0$. By trans\-forming to the Majoranas (Greek letters) ,  $\beta_{i} \!=\! (c_{i}+c_{i}^{\dagger})/\sqrt{2}$ and $\gamma_{i} \!=\! (-i c_{i}+i c_{i}^{\dagger})/\sqrt{2}$, our model reduces to two decoupled Majorana chains, as shown in Fig.\,\ref{fig:setup}(b). The side-coupled MBS in the lower chain corresponds to the MBS $\eta_1$. The conductance through the dot is, then, the sum of the conductance from two decoupled Majorana chain $G=G^{\textrm{upper}}+G^{\textrm{lower}}$.

Consider now two other cases. First, for a system without a side-coupled mode, the Majorana representation leads to two decoupled chains as shown in Fig.\,\ref{fig:setup}(c). Second, for a system with a side-coupled regular fermionic zero mode, the Majorana representation consists of two decoupled chains, each of which has a side-coupled MBS [Fig.\,\ref{fig:setup}(d)]. For both cases, $H^{\textrm{upper}} \!=\! -H^{\textrm{lower}}$, and thus $G^{\textrm{upper}} \!=\! G^{\textrm{lower}}$. Since the peak conductance for a dot with (without) a side-coupled regular fermionic zero mode is 0 ($e^2/h$), the result for a single Majorana chain with (without) a side-coupled MBS is 0 ($e^2/2h$).Therefore, the conductance of our model [Fig.\,\ref{fig:setup}(b)] is $G_\textrm{peak} \!=\! 0+e^2/2h=e^2/2h$.  

The spectral function of the dot, $A(\omega)= -2\Gamma \operatorname{Im}[G_{dd}^R(\omega)]$, is shown in Fig.\,\ref{fig:MBS_spectral}(a) for several values of the dot-MBS coupling $\lambda$ and dot-lead coupling $\Gamma$ for $\epsilon_M=0$. The energy unit is chosen so that the lead band width is $D_L=40$ throughout the paper. Consistent with our assumption that the Zeeman splitting is the largest energy scale, we consider the spectrum for only the spin-down channel. For $\lambda=0$, the spectral function reduces to the result of the
resonant level model. For small dot-MBS coupling %$\lambda$ 
($\lambda=0.02, 0.05$), % and $\Gamma=0.2$), 
the spectrum shows two peaks at $\omega\sim \pm \lambda$ which come from the energy level splitting caused by coupling to the MBS. As we increase $\lambda$ with fixed $\Gamma=0.2$, the two peak structure evolves into a spectrum with three peaks, showing clearly the presence of the Majorana zero mode. Note that the zero frequency spectral function always gives $A(\omega=0)=1/2$ as long as $\epsilon_M=0$ and $\lambda\neq 0$. For small dot-MBS coupling ($\lambda=0.02$), the three peak spectrum also appears upon decreasing $\Gamma$.

The dot spectrum for different strengths of MBS-MBS coupling $\epsilon_M$ appears in Fig.\,\ref{fig:MBS_spectral}(b). Even for very small coupling $\epsilon_M=0.02$, the zero frequency spectrum shows $A(\omega=0)=1$ not $1/2$. The width of the narrow peak is proportional to $\epsilon_M$. For large coupling ($\epsilon_M=0.3$), the spectrum reduces to the resonant level result along with two additional small peaks at $\omega\sim\pm\epsilon_M$ corresponding to the energy of the effective Dirac fermionic state $f$. If the wire is long enough so that $\epsilon_M\ll T, \lambda$, one can still observe the $G_\textrm{peak} \!=\! e^2/2h$ signature.

{\em More Realistic Wire---}To analyze the robustness of the 
%$e^2/2h$ 
MBS signature in the real physical system, the single MBS in 
Eq.\,(\ref{eq:H_DM})  is replaced by the whole nanowire \cite{lutchynPRL10,oreg10} shown in Fig.\,\ref{fig:setup}(a). 
We study numerically a lattice tight-binding Hamiltonian \cite{potterPRB11},
$H_{\rm{wire}} \!=\! H_{0}+H_{\rm{Rashba}}+H_{\rm{SC}} $,
where $H_0$ includes nearest-neighbor hopping along the wire ($\hat{y}$ direction), a chemical potential leading to half filling ($\mu=0$), and a magnetic field perpendicular to the surface ($\hat{z}$ direction) causing the Zeeman splitting $V_z$. 
The Rashba spin-orbit interaction is
\begin{equation}
H_{\rm{Rashba}}=\sum_{i,s s^{'}}-i\,\alpha_{R}w_{i+1,s}^{\dagger}\hat{z}\cdot(\vec{\sigma}_{s s^{'}}\times\hat{x})\, w_{i,s^{'}}+h.c.
\end{equation}
%The Rashba interaction $\alpha_{R}$ is along $\hat{x}$ direction. 
% HUB: \alpha_R is a scalar, so it can't have a direction...
where $w_{i,s}^{\dagger}$ creates an electron with spin index $s$ on site $i$ of the wire and 
$\overrightarrow{\sigma}$ are the Pauli matrices. 
Finally, the \emph{s}-wave pairing term with superconducting order parameter $\Delta$ is
\begin{equation}
H_{\rm{SC}}=\Delta\sum_{i}w_{i,\uparrow}^{\dagger}w_{i,\downarrow}^{\dagger}+h.c.
\end{equation}
The Bogoliubov-deGennes equation is constructed from $H_\textrm{wire}$ by the standard Nambu spinor 
representation (including the same Zeeman splitting $V_z$ in the dot) and then solved by a recursive Green function method \cite{FerryGoodnickBird,asano01}. 
% HUB: look for more recent ref with the BdG done, or use a textbook. 

\begin{figure}[t]
\centering
\includegraphics[width=3.3in,clip]{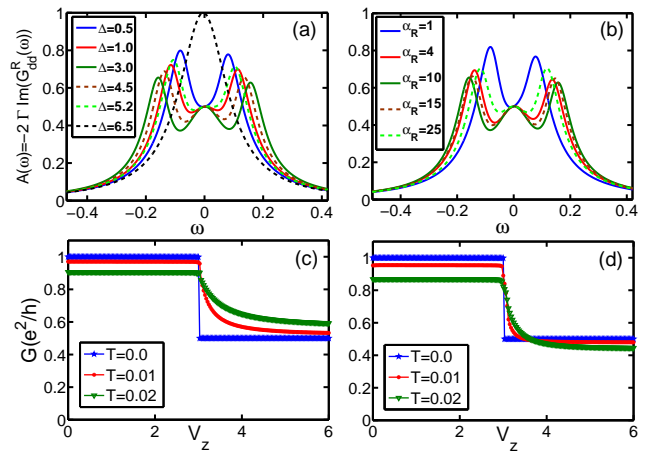}
\caption{(color online) Dot spectral function and peak conductance in the more realistic nanowire case (the dot is on-resonance and symmetrically coupled to the probe leads). $A(\omega)$ for different values of (a) the SC order parameter at fixed $\alpha_R \!=\! 2$ and (b) the Rashba interaction strength at fixed $\Delta \!=\! 3$. The results are qualitatively similar to those of the simple model (Fig.\,2). (Parameters: $\Gamma \!=\! 0.1$, $\lambda \!=\! 0.3$, and $V_z \!=\! 6$.)
(c),(d) Conductance as a function of Zeeman energy for different temperatures at fixed $\Delta \!=\! 3$. The sharp change at $V_z \!=\! \Delta$ is a signature of the topological phase transition.
(Parameters: (c) $\alpha_R \!=\! 2$, $\lambda \!=\! 0.1$, $\Gamma \!=\! 0.1$; 
(d) $\alpha_R \!=\! 10$, $\lambda \!=\! 0.3$, $\Gamma \!=\! 0.08$.) Throughout, $\mu \!=\! 0$, the hopping in the nanowire $t \!=\! 10$ corresponds to a band width $D \!=\! 40$, and the wire consists of $1000$ sites.
} 
\vspace{0.1in}
\label{fig:wire_spectral}
\end{figure}

The dot spectral function is shown in Fig.\,\ref{fig:wire_spectral} for several values of the SC order parameter $\Delta$ and Rashba interaction strength $\alpha_R$ (for an on-resonance, symmetrically coupled dot) \cite{exptvalues}. When the wire is in the topologically trivial phase ($\Delta \!>\! V_z$, no MBS), the spectrum is similar to the resonant level result [Fig.\,\ref{fig:wire_spectral}(a)]. In contrast, when the wire is in the topological SC phase ($\Delta \!<\! V_z \!=\! 6$, $\mu \!=\! 0$), the value of the spectral function at zero 
frequency is $1/2$. For $\Delta$ small ($\Delta \!=\! 0.5$), the spectrum shows
two peaks, but upon increasing $\Delta$ ($\Delta \!=\! 1, 3$), the two peaks become more separate and the three-peak structure emerges. Curiously, a further increase of $\Delta$ ($\Delta \!=\! 4.5, 5.2$) leads to a smaller separation between the outer peaks. Similar phenomena occur upon varying the Rashba interaction $\alpha_R$: increasing $\alpha_R$ leads to first an increase in the splitting of the outer peaks ($\alpha_R \!=\! 1, 4, 10$) and then a decrease ($\alpha_R \!=\! 15, 25$). 

The non-monotonic shifts in the positions of the outer peaks can be understood as follows. When $\Delta$ or $\alpha_R$ is small, the \emph{p}-wave SC pairing $f_{p}$ 
is weak, leading to a less robust MBS and small peak splitting. 
On the other hand, when $\Delta$ is large and close to the transition value $\Delta \!=\! V_z$, 
SC pairing between the lower and upper band \cite{aliceaPRB10} makes the MBS 
less robust. For large $\alpha_R$, the eigenfunction of the lower band at the fermi surface has a large spin-up component, while the dot and leads are spin-down due to the Zeeman splitting; 
therefore, the coupling between the dot and MBS is suppressed. As a function of both parameters, then, there is non-monotonic behavior.

To detect the MBS, a clear signature appears in the conductance as a function of Zeeman splitting [Fig.\,\ref{fig:wire_spectral}(c) and (d)]: the conductance at zero temperature shows a sharp jump at $V_z \!=\! \Delta$ due to a topological phase transition \cite{exptvalues}. For $V_z<\Delta$, the wire is in the topologically trivial phase, and the peak conductance is $e^2/h$. For $V_z>\Delta$, the wire is in the topological SC phase in which a MBS appears, and the peak conductance is $e^2/2h$. (Both of these values are multiplied by the factor $4\Gamma_L \Gamma_R/(\Gamma_L + \Gamma_R)^2$ for asymmetric coupling to the leads.) At finite temperature, the jump becomes a crossover, which is still quite sharp near the transition point. For small $\alpha_R$, $\lambda$ and large $\Gamma$, the spectrum has two peaks, so the finite $T$ conductance is larger than $e^2/2h$ [Fig.\,\ref{fig:wire_spectral}(c)]. For large $\alpha_R$, $\lambda$ and small $\Gamma$, the spectrum has three peaks, causing the finite $T$ conductance to be smaller than $e^2/2h$ [Fig.\,\ref{fig:wire_spectral}(d)]. 

We emphasize that the change in conductance by a factor of $1/2$ is universal as long as the MBS appears and couples 
to the dot. With regard to the effect of disorder in the wire \cite{lutchyn11}, a short range impurity potential does 
not affect the MBS and thus the $G_{\rm peak} \!=\! e^2/2h$ result, while a sufficiently strong long range impurity potential may 
induce mixing of the MBS at the two ends and therefore lead to $G_{\rm peak} \!=\! e^2/h$ as shown in Fig.\,\ref{fig:MBS_spectral}(b).

\begin{figure}[t]
\centering
\includegraphics[width=3.3in,clip]{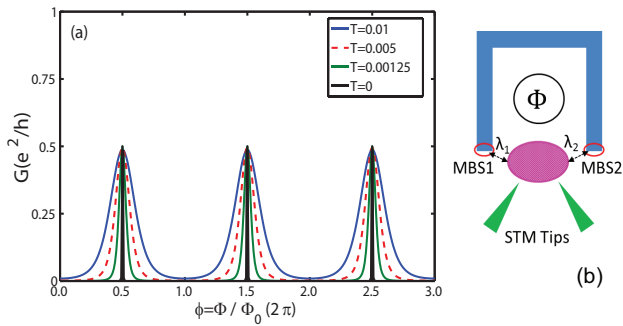}
\caption{(color online) (a) Conductance for MBS-dot-MBS system as a function of 
the phase $\phi=\Phi/\Phi_0$ for different temperatures; the dot is on-resonance and symmetrically coupled to the STM tips.
($T=0.01$, $0.005$, $0.00125$, and $0$, from top to bottom; parameters are
$\lambda_1=\lambda_2=\Gamma_1=\Gamma_2=0.1$.) This curve does not depend on the value of $|\lambda_1/\lambda_2|$.
(b) Sketch of MBS-dot-MBS system. The two MBS appear at the ends of the nanowire;
$\Phi$ is the magnetic flux through the loop. The conductance is measured using dual-tip STM, allowing tuning to the degeneracy point.
} 
\vspace{0.1in}
\label{fig:twoMBS}
\end{figure}

{\em Two MBS---} Consider the geometry proposed by Flensberg \cite{flensberg11} for implementing a $\pi/8$ phase gate: a dot coupled to both ends of the nanowire---and hence to two MBS---with 
magnetic flux $\Phi$ through the loop, as shown in Fig.\,\ref{fig:twoMBS}(b). The conductance through the dot is measured using two external leads; since electron tunneling between the dot and environment should be avoided during qubit operation, a dual-tip STM setup \cite{yi05,jaschinsky08} is proposed so that one can remove the external leads after tuning the system.
The Hamiltonian of
this MBS-dot-MBS system \cite{flensberg11} can be written as
\begin{equation}
H=\epsilon_d d^{\dagger}d+(\lambda_{1}^{*}d^{\dagger}-\lambda_{1}d)\eta_{1}+i(\lambda_{2}^{*}d^{\dagger}+\lambda_{2}d)\eta_{2} \;.
\end{equation}
The phase difference between the two couplings, $\phi \!\equiv\! 2\arg(\lambda_{1}/\lambda_{2})$, is related to the flux 
$\Phi$ via $\phi\!=\!\Phi/\Phi_0$, where $\Phi_0\!=\!h/2e$. 
Without loss of generality, we take $\lambda_1$ to be real ($\lambda_1\!=\!|\lambda_1|$ and $\lambda_2\!=\!|\lambda_2|e^{-i\phi/2}$),
and the Hamiltonian reduces to
% \begin{equation}
%  H  =  \epsilon d^{\dagger} d + \lambda (d^{\dagger} \eta_{12} +\eta_{12}^{\dagger} d)
% \end{equation}
$H  \!=\!  \epsilon d^{\dagger} d + \lambda (d^{\dagger} \eta_{12} +\eta_{12}^{\dagger} d)$
where $\eta_{12} \!\equiv\! (|\lambda_1|\eta_1+i e^{i\phi/2}|\lambda_2|\eta_2)/\lambda$ and 
$\lambda \!\equiv\! \sqrt{|\lambda_{1}|^{2}+|\lambda_{2}|^{2}}$. For $\phi\!=\!(2n+1)\pi$ ($n$ integer),
we have  $\eta_{12} = \eta_{12}^{\dagger}$. In this case, the dot is effectively coupled to a single MBS $\eta_{12}$;
therefore, the $T=0$ on-resonance conductance is $e^2/2h$. For $\phi\neq (2n+1)\pi$, we have
$\eta_{12}\neq \eta_{12}^{\dagger}$ corresponding to a regular fermionic zero mode, for which the $T=0$ on-resonance
conductance is zero.

Following the method for single MBS, one can exactly solve for the dot Green function $G_{dd}^R(\omega)$ in this
two MBS problem in the case of $\epsilon_M=0$:
\begin{equation}
% G_{dd}^{R}(\omega)=\frac{1}{\left[G_{dd}^{R0}(\omega)\right]^{-1}-A(\omega)-B(\omega)}
G_{dd}^{R}(\omega)= \left\{ \left[G_{dd}^{R0}(\omega)\right]^{-1}-A(\omega)-B(\omega) \right\}^{-1}
\end{equation}
where $A(\omega)=-i \Gamma + (|\lambda_1|^2+|\lambda_2|^2) / 2\omega$ and 
\begin{equation}
 B(\omega)=\frac{\frac{1}{4 \omega^2} \big[|\lambda_1|^4+|\lambda_2|^4+2|\lambda_1|^2 |\lambda_2|^2 \cos(\phi-\pi) \big]}{\omega+\epsilon_d+i \Gamma-(|\lambda_1|^2+|\lambda_2|^2) / 2\omega} \;.
\end{equation}

The conductance peak value as a function of the phase difference $\phi$ can be obtained from  Eq.\,(\ref{KuboFormular}) and is shown 
in Fig.\,\ref{fig:twoMBS}(a). For $T=0$, the $G=e^2/2h$ signature appears 
only at $\phi=(2n+1)\pi$, 
%($n$ integer)
corresponding to the energetically degenerate state in Flensberg's qubit \cite{flensberg11}, with $G=0$ otherwise. For $T\neq 0$, 
the peak width becomes finite; note that the peak is fairly wide even for $T=0.01$ but that the temperature is still low enough to 
see the MBS. By tuning the conductance to a resonance peak, one can tune the MBS-dot-MBS to the desired degenerate energy point.

%In conclusion, we propose a way to probe the presence of a MBS by using the conductance through a quantum dot: the signature is a conductance of $e^2/2h$ in contrast to either 0 (regular fermionic zero mode) or $e^2/h$ (no zero mode). 
%In our proposal, electron transfer between the MBS and outside system 
%is not necessary, though dephasing of the qubit is introduced when the dot is on-resonance. Such a ``less invasive'' sensing method provides a potential way 
%to probe a MBS without totally destroying the information in the qubit (the readout problem); still, most probably, error correction will be needed. In addition, we provide a way to tune Flensberg's 
%topologically unprotected $\pi/8$ phase gate to the degeneracy point. 

\medskip
This work was supported by the U.S.\,DOE (Office of Basic Energy Sciences, Division of Materials Sciences and Engineering, award DE-SC0005237) and the U.S.\,Office of Naval Research (two MBS and quantum computing). HUB appreciates the hospitality of the Fondation Nanosciences in Grenoble, France, during the completion of this work. 

%\vspace*{-0.25in}
\bibliography{Majorana}

\begin{thebibliography}{27}
\bibitem[{num()}]{numberraiselower}
\bibinfo{note}{Strictly speaking, the last two terms in $H$ should include
  raising and lowering operators for the number of Cooper pairs in the
  ungrounded, bulk superconductor. However, using the BCS wavefunction, we can
  show that these operators simply introduce a phase factor which can then be
  absorbed into $d$ and $f$.}
\bibitem[{exp()}]{exptvalues}
\bibinfo{note}{The parameters used in Fig.\,3 fall within the possible range of
  parameters for an InAs wire on Nb. For typical experimental values taken from
  Ref.\,\onlinecite{lutchynPRL10} and references therein, and for $B_z$ about
  half of the critical field of Nb, the parameters used in Fig. 3(c),
  $(\Delta_{SO}, V_z, \Delta)=(11,6,3)$, correspond to $(220,120,60)\mu$eV
  while the experimental values are $(\sim\!250, \sim\!120,
  \sim\!30$-$120)\mu$eV.}
\end{thebibliography}

\end{document}